\documentclass[aps,prl,twocolumn,10pt,superscriptaddress]{revtex4-2}

\usepackage{amsmath,amssymb,graphicx,bm}
\usepackage{times}
\usepackage{lipsum}
\usepackage{hyperref}
\usepackage{enumitem}

\begin{document}

\title{Copy–Spread–Annihilate Dynamics in Degree-Assortative Networks}

\author{Yan Hao }
\email{hao@hws.edu}
\affiliation{Hobart and William Smith Colleges, Geneva, NY, USA}
\author{Daniel J. Graham}
\affiliation{Hobart and William Smith Colleges, Geneva, NY, USA}
\author{Marc-Thorsten Hütt}
\affiliation{Constructor University, Bremen, Germany}


\date{\today}

\begin{abstract}
In many systems, communication proceeds by broadcasting rather than single source–target routing, but network structures that maximize signal lifetime are not well understood. Degree correlations are known to influence robustness and spreading, yet their effect on signal persistence has remained unclear. Here we introduce Copy–Spread–Annihilate dynamics, a minimal synchronous broadcasting model with annihilation. We show that signal lifetimes vary non-monotonically with assortativity and are maximized near neutral assortativity, where hub-driven amplification is strong but annihilation via short cycles is still limited. Applying this framework to the mouse connectome suggests assortativity as a structural control parameter for broadcast signal persistence in brain-like and other complex networks.
\end{abstract}

\maketitle
\section{Introduction}
\begin{figure}
\centering
    \includegraphics[width=0.75\linewidth]{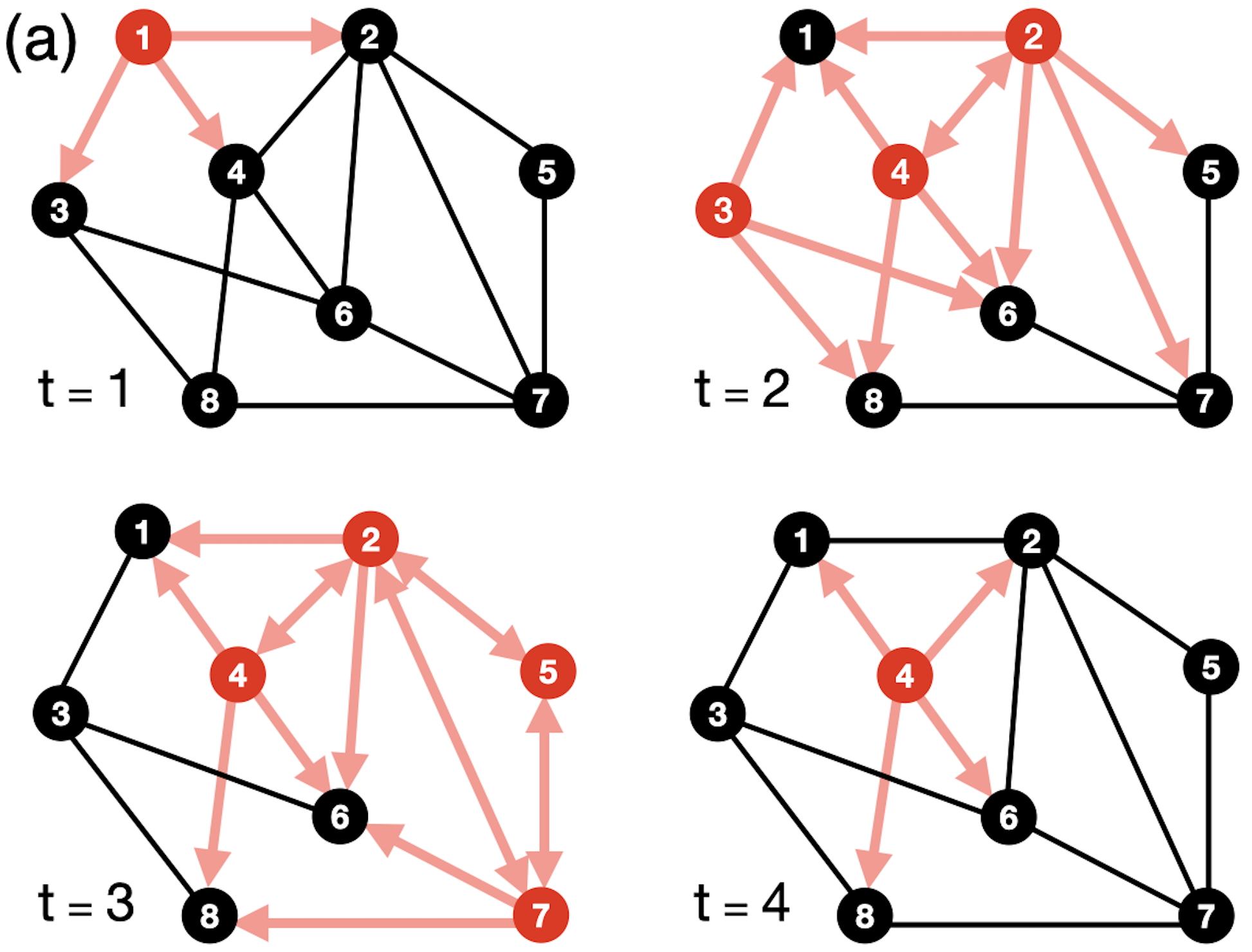}
    \includegraphics[width=0.9\linewidth]{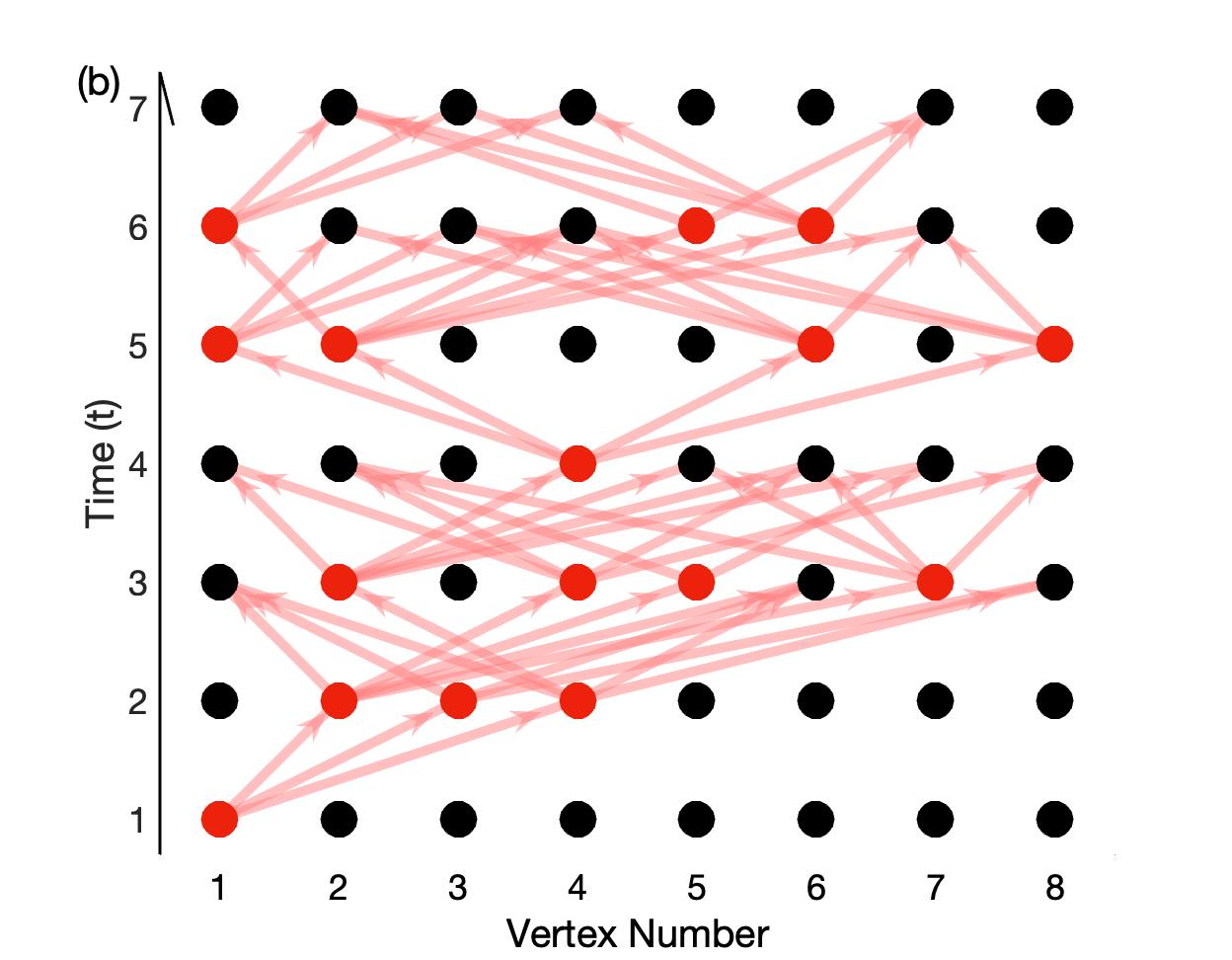}
    \caption{Visualization of Copy-Spread-Annihilate dynamics implemented on a small network, shown in the single injection case for clarity. (a) demonstrates, frame-by-frame, how one message (injected to vertex 1) spreads, and how copies resulting from this single injection interfere with each other. (b) plots the node space versus time (space-time diagram). It illustrates the propagation of the same message and its copies through its lifetime until full self-annihilation. In this particular case, the initial message created 9 different walks that achieved the maximum distance $\tau_M = 6$. }
    \label{fig:scheme}
\end{figure}

How long can a message live as it spreads across a complex network? This question is central to both information dynamics and biological signaling. Here we address it with a simple synchronous rule we call \emph{Copy--Spread--Annihilate Dynamics}: at each discrete step, every vertex containing a message copies its message and sends (i.e., broadcasts) them to all neighbors; vertices receiving exactly one copy retain the message to the next step, while two or more incoming messages annihilate each other (Fig. \ref{fig:scheme}). New messages are injected each timestep at a fraction of unoccupied vertices. The quantity of interest is the \emph{message lifetime}, defined as the time until the last copy of an injected message disappears.

Network communication is an important and growing area of investigation \cite{newman2023message}. The study of broadcasting-like behavior on networks has long been approached from a graph theory perspective \cite{bavelas1950communication, landau1954distribution, shimbel1951applications, hajnal1972cure, hedetniemi1988survey, harutyunyan2017broadcast}. Dynamical systems theory has similarly investigated spreading behaviors such as epidemics and reaction-diffusion processes \cite{ji2023signal}. Unlike specific routing, which delivers information from a source to a target without disturbing the rest of the system (at least in idealized systems) \cite{trusina2005communication}, broadcasting amplifies messages by duplicating them across multiple outgoing edges, resembling worms \cite{wang2013modeling}, blockchain communication \cite{narayanan2016bitcoin}, or viral spread \cite{trusina2004hierarchy}. 

However, message spread can be inhibited by messages themselves, as has been observed on social networks \cite{liang2017information, liang2019network}. This behavior highlights the importance of understanding the impact of message-message interactions on message dynamics in broadcasting systems, which have so far not been investigated.

In the brain, the global neuronal workspace has been hypothesized as a broadcasting system, ensuring simultaneous access to information across a distributed communication network \cite{baars1997theatre, wajnerman2022global, fakhar2025general, zeki1998asynchrony, zeki2020multiplexing}. This idea stands in contrast to specific routing models of the brain which range from shortest-path navigation to random walk models \cite{avena2019spectrum}. While specific routing models are widely used, few models consider message-message interactions, which could compromise communication efficiency in a host of ways \cite{graham2014routing,graham2020network,graham2023nine}. These interactions could include classical inhibition, as well as XOR or other gating mechanisms \cite{steriade2007gating, gollisch2010eye, gidon2020dendritic, oz2022non}. Queueing models do attempt to manage message interactions \cite{misic2014communication,misic2014network, fukushima2024packetization}, but these models invoke unrealistic message addressing and caching mechanisms. 

Our Copy–Spread–Annihilate model \cite{hao2020creative} defines a minimal Markovian broadcasting scheme where collisions serve as both coincidence-detection events and annihilation events, providing a parsimonious model of how message lifetime depends on network structure. Though the CSA model is not designed for efficiency, it provides an ideal testbed for identifying which purely structural features of a network enhance or suppress the lifetime of spreading messages in the presence of message-message interactions.

A central structural parameter in network science is \emph{degree assortativity}, the correlation between degrees of adjacent vertices \cite{newman2002assortative,newman2003mixing}. Positive assortativity indicates hubs connected to hubs; negative assortativity indicates hub-leaf connectivity. Prior work has shown that assortativity influences robustness, percolation thresholds, and spreading dynamics \cite{boguna2003epidemic, vazquez2003resilience, newman2002assortative}. Yet the impact of assortativity on lifetimes of self-annihilating messages has not been studied.

We find that message lifetime exhibits a \emph{non-monotonic} dependence on assortativity: lifetimes peak near neutral assortativity and decrease at high positive and negative values. This result is counterintuitive, as one might expect monotonic effects of assortativity on spreading capacity. Our analysis shows that the peak arises from a balance between two opposing mechanisms: amplification and self-annihilation of messages. We show that the structural ingredients that promote maximal message lifetime are reflected in the mammal connectome, and we propose that this result helps explain why such networks have the evolved structure they do.
\section{Model and Methods}

\paragraph{Copy--Spread--Annihilate (CSA) dynamics.} Let $G=(V,E)$ be a finite, simple, undirected graph with vertex set $V$ and edge set $E$. Messages are introduced to the graph using either of the following methods: {\bf single injection}, in which one message is placed on a vertex and is allowed to spread without additional injections; {\bf sequential injection}, at each time step, one new message is injected into an unoccupied vertex on the graph ({\footnotesize see S1. and Fig. \ref{fig:chart1} for numerical implementation details}).  \\
The dynamical state of the vertices at discrete time $t \in \mathbb{N}$ is given by a configuration 
$X_t : V \to \{0,1\}$, where $X_t(v)=1$ if vertex $v$ is occupied by a message copy at time $t$ 
and $X_t(v)=0$ otherwise. For each vertex $v$, denote by $N_G(v)$ its set of neighbors. 
The update rule is synchronous and defined as:
\begin{equation}[floatfix]
X_{t+1}(v) = \mathbf{1}_{\left\{\sum_{u\in N_G(v)} X_t(u) = 1\right\}}(v),
\label{eq:update}
\end{equation}
where $\mathbf{1}_A(v)$ denotes the indicator function of set A. 
Thus, each occupied vertex spread copies to all neighbors; a vertex is occupied at the next step if and only if it receives exactly one inbound copy. 
Although this paper focuses on sequential injection, Fig. \ref{fig:scheme} demonstrates the CSA dynamics using the single injection protocol for clarity. For schematic diagrams of the sequential injection case, see Figures \ref{fig:PathLow},\ref{fig:PathMid}, and \ref{fig:PathHigh}.




\paragraph{Message lifetime.}
Any message, $M$, entering the graph 
via the initial source vertex $v_M \in V$ creates many walks during its passage through the graph due to spreading. For example, in Fig. \ref{fig:scheme}(b), the message created 43 walks during its lifetime. In general, the $i$th walk traversed by a copy of $M$ can be described by the vertex sequence
\begin{equation}
P_{v_M,i} = \left(v_M,v_{i_1},v_{i_2},\dots,v_{i_s} \right),
\label{eq:path}
\end{equation}
$v_{i_s}$ being the vertex where this copy is annihilated. The distance of this walk is denoted $D_{v_M,i} =s $. 
The \emph{message lifetime} $\tau_M$ is the maximum distance of any walk in the broadcast generated by message $M$:
\begin{equation}
\tau_M = \max \{\, D_{v_M,i} \in \mathbb{N} \,\}.
\label{eq:lifetime}
\end{equation}
Let $\mathcal{P}_{v_M}$ denote the set of all longest walks realized by occupied vertices up to time $\tau_M$. 
\begin{equation}
  \mathcal{P}_{v_M} = \{P_{v_M,i}: D_{v_M,i} = \tau_M\}.
  \end{equation}
For each graph $G$, we report $\langle \tau \rangle$, the average of $\tau_M$ weighted by $|\mathcal{P}_{v_M}|$, i.e. the number of walks that reached $\tau_M$, over all messages injected to the graph.
\vspace{0.1in}
\paragraph{Assortativity.}
The degree assortativity coefficient, $r$, is the Pearson correlation coefficient of the degrees at the ends of edges \cite{newman2002assortative,newman2003mixing}:
\begin{equation}
r = \frac{\sum_{(u,v)\in E} (k_u - \bar k)(k_v - \bar k)}
         {\sum_{(u,v)\in E} (k_v - \bar k)^2},
\end{equation}
where $k_v$ is the degree of vertex $v$, and $\bar k$ is the mean degree of all vertices. 
\vspace{0.13in}
\paragraph{Amplification and self-annihilation proxies.}
We quantify two structural drivers of message lifetime:
\begin{enumerate}
  \item \emph{Amplification} is captured by ``hub neighbor degree," i.e., the average degree of hub vertices' neighbors: 
  \begin{equation}
  k_{\text{hn}}(k^*) = \frac{1}{|\{v : k_v>k^*\}|} \sum_{v : k_v>k^*} \left(\frac{1}{k_v} \sum_{u\in N_G(v)} k_u\right),
  \end{equation}
where a ``hub" refers to any vertex whose degree is ranked above the 90th percentile, and thus $k^* = P_{90}$. This quantity reflects how strongly hubs broadcast to other well-connected vertices. Intuitively, high $k_{\text{hn}}(k^*)$ values indicate that hubs preferentially
connect to other hubs, thus messages will be broadcast by a hub, and then amplified by other hubs on the next timestep. Low $k_{\text{hn}}(k^*)$ values indicate that hubs connect to low-degree vertices which cannot efficiently propagate messages, thus reducing amplification.
  \item \emph{Self-annihilation} is captured by the number of 4-cycles in the graph. They provide two equal-length paths that can deliver simultaneous arrivals of the same message and thus self-annihilation. Conceptually, the number of 4-cycles in any graph is
    \begin{equation}
  C_4(G) = \frac{1}{8} \sum_{(v_i,v_j,v_k,v_{\ell})} \mathbf{1}\{(v_i,v_j),(v_j,v_k),(v_k,v_{\ell}),(v_{\ell},v_i) \in E\},
  \end{equation}
where $(i,j,k,l)$ are distinct ordered quadruples. The factor $\tfrac{1}{8}$ accounts for symmetries. $C_4$ can be approximated using the adjacency matrix $A$:
  \begin{equation}
  C_4(G) \approx \frac{1}{8}\left( \mbox{Tr}(A^4) -2\sum_{i=1}^{|V|} k_{v_i}^2 + \sum_{i=1}^{|V|} k_{v_i} \right).
  \end{equation}
  
\end{enumerate}

\paragraph{Vertex participation in longest walks.}
For a vertex $v$, we define the participation probability
\begin{equation}
p(v) = \mathbb{P}\{ v \in P \mid P \in \mathcal{P}_{v_M}\},
\end{equation}
conditioned on the sampled sources $v_M$. We then examine histograms of $p(v)$ across all vertices to quantify how uniformly or selectively vertices contribute to sustaining longest message lifetimes under different assortativity regimes.

\section{Results}
\begin{figure}
\centering
    \includegraphics[width=0.9\linewidth]{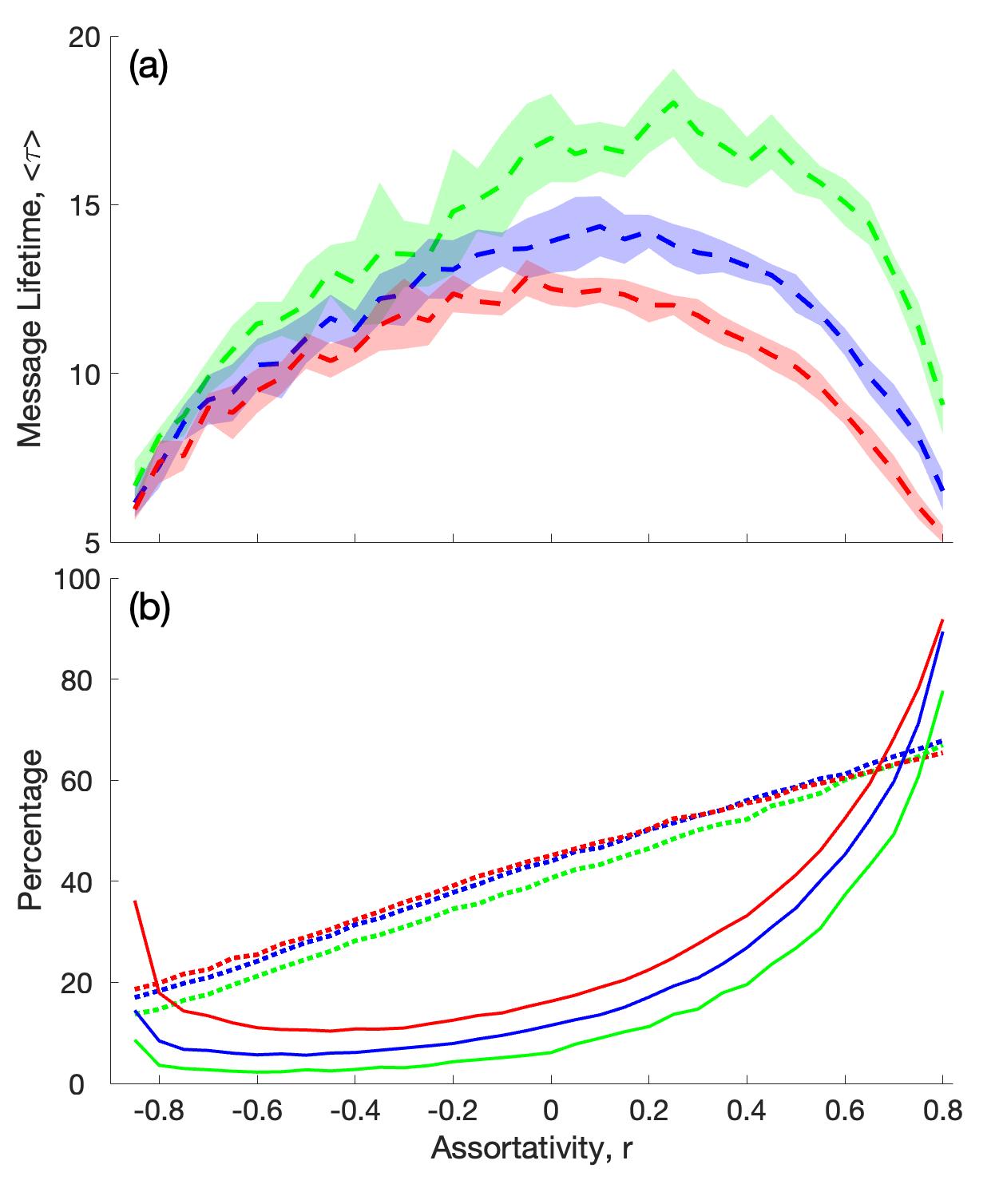}
    \caption{(a) Average message lifetime as a function of assortativity shown in dashed curves. Shaded areas indicate standard deviations. Results are collected from five different networks of 100 vertices, generated using Barabási-Albert (BA) algorithm, and $m$ = 2, 3, 4 as parameters of attachment (in green, blue, and red respectively).  The spectrum of assortativities are achieved using a degree-based edge-switching algorithm \cite{trusina2004hierarchy} in order to preserve degree sequence.  (b.) Using the same graphs as in (a), the average neighbors' degrees of hub vertices, $k_{hn}$, and the number of 4-cycles, $C_4(G)$, are presented in dotted and solid curves, respectively. See supplementary material Sec. S1 for simulation details. }
    \label{fig:LifeTime1}
\end{figure}

\begin{figure}
\centering
    \includegraphics[width=0.8\linewidth]{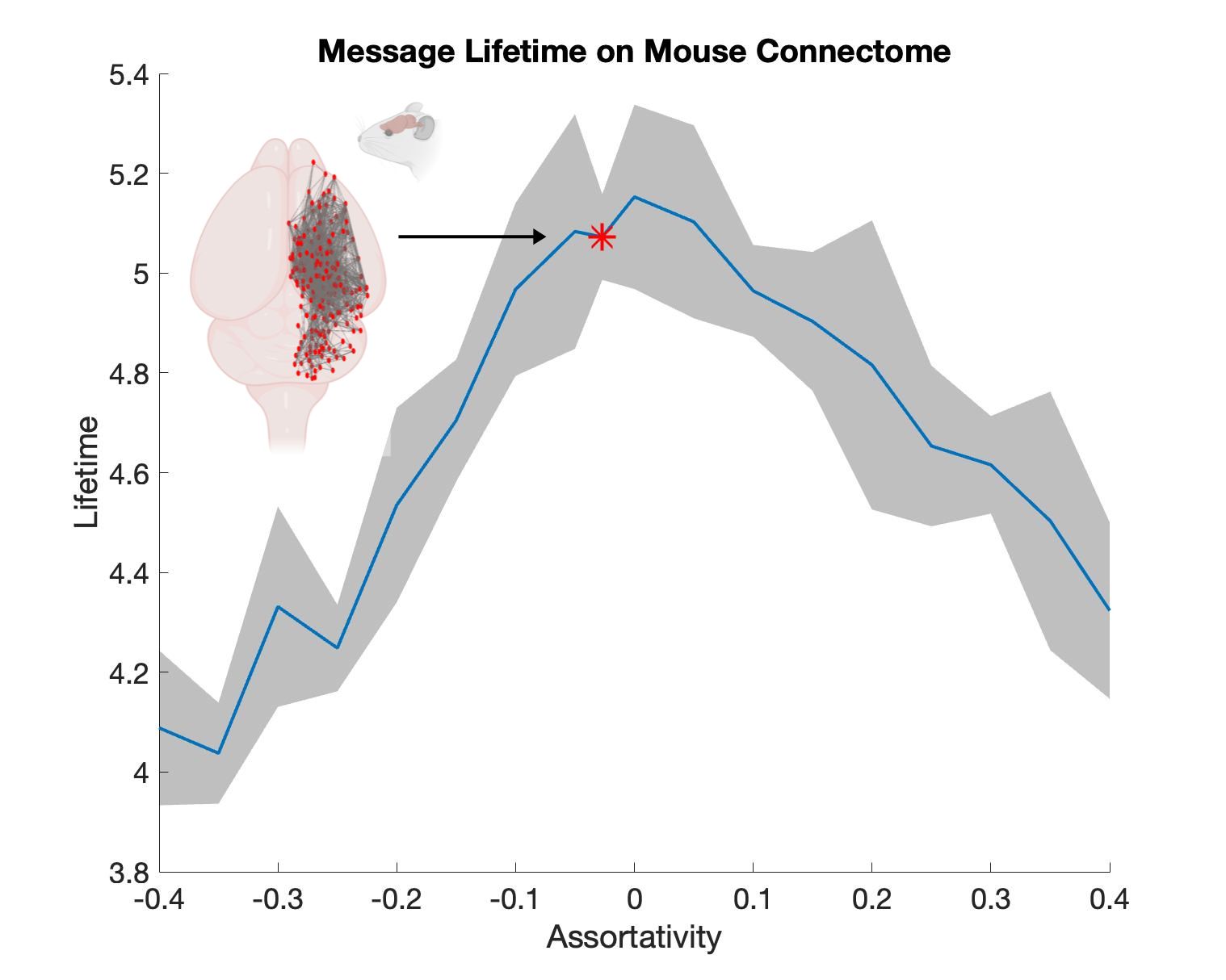}
    \caption{Average message lifetime as a function of assortativity on empirical mouse connectome (asterisk) \cite{oh2014mesoscale} and surrogate graphs (curve). Five sets of surrogate graphs are generated using the same edge-switching algorithm as in Fig. \ref{fig:LifeTime1} to achieve a spectrum of assortativities. The shaded area indicates standard deviations across five repetitions of all graphs.}
    \label{fig:mouse}
\end{figure}

\paragraph{Non-monotonic lifetimes in BA networks.}
Fig. \ref{fig:LifeTime1}(a) shows the average message lifetime as a function of assortativity in Barabási–Albert graphs of 100 vertices. The curve is robustly non-monotonic across different densities, peaking near neutral assortativity ($r$ close to zero). Variance across five sets of different graphs and five repeated simulations on each set of graphs are demonstrated by the shaded area. The non-monotonic behavior is also independent of the size of the graphs and the number of messages that are injected at each time step (Fig. \ref{fig:LifeTime2}).
\vspace{0.05in}
\paragraph{Mechanistic decomposition.} 
In Fig. \ref{fig:LifeTime1}(b) we compare normalized values of average hub neighbor degrees and 4-cycle counts as functions of assortativity. Across the assortativity spectrum, amplification increases monotonically while self-annihilation first decreases, then increases. Maximal lifetimes occur where amplification is high but self-annihilation is still limited, i.e.\ near neutral assortativity. Annihilation caused by copies from different messages is not a primary condition that contributes to the non-monotonic shape of the lifetime
graphs ({\footnotesize see supplementary material Sec. S4 for details}).
\begin{figure}
    \centering
    \includegraphics[width=0.95\linewidth]{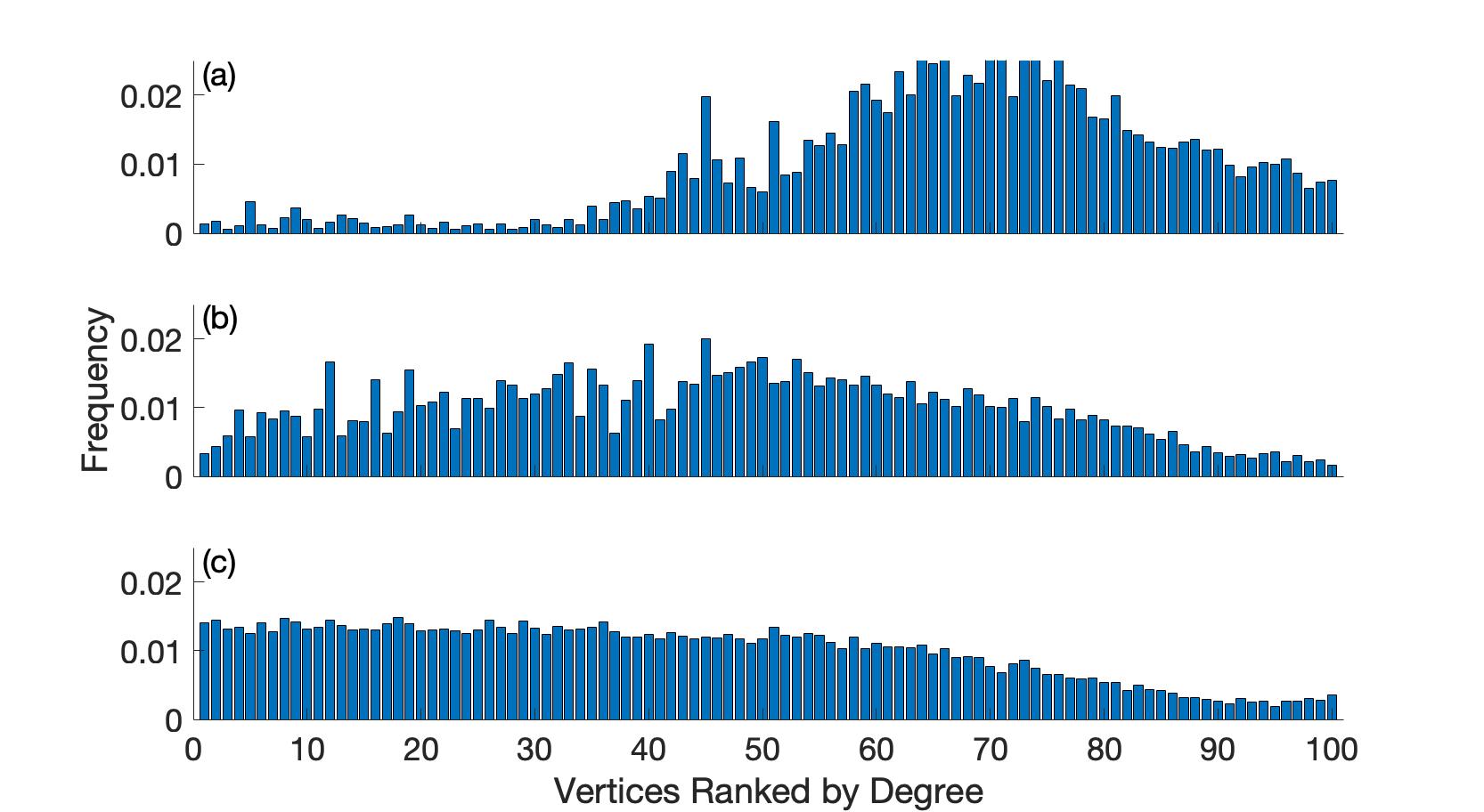}
    \caption{Vertex participation frequency in walks that are above the 90th percentile in length, produced using five different BA graphs with 100 vertices and $m$ = 3. Vertices are sorted by degree. (a) high-degree vertices are preferred in disassortative graphs ($-0.85<r< -0.65$), (b) shows the vertex usage when assortativities are neutral ($-0.1<r< 0.1$), (c) low-degree vertices are preferred in assorted graphs ($0.6<r<0.8$). See supplementary material Sec. S1 for simulation details.}
    \label{fig:NodeUsage}
\end{figure}
\vspace{0.05in}
\paragraph{Vertex participation.}
Upon analyzing histograms of vertex participation in longest walks, Fig. \ref{fig:NodeUsage} reveals distinct strategies for sustaining messages:
\renewcommand{\theenumi}{\roman{enumi}}
\begin{enumerate}[noitemsep]
\vspace{-0.05in}
    \item  at high negative assortativity histograms skewed to high-degree vertices implying successful walks tend to involve high-degree vertices as amplification is concentrated through hubs;
    \item at near neutral assortativity, all vertices participate roughly equally; 
    \item at high positive assortativity, long-lived messages travel mainly via low-to-medium-degree vertices, as collisions in hub cores prune those vertices from successful walks, shifting participation to lower-degree vertices.
\end{enumerate}
\vspace{-0.05in}
Successful walk examples from smaller graphs also support these observations ({\footnotesize see Fig. \ref{fig:PathLow},\ref{fig:PathMid},\ref{fig:PathHigh}}). 
\vspace{0.05in}
\paragraph{Neuroscience relevance.}
Finally, we test the model on an empirical mammalian connectome and its surrogates (Fig \ref{fig:mouse}). The mouse connectome \cite{oh2014mesoscale} was converted into an undirected graph by retaining all two-way edges that have non-zero weights, and setting all weights to 1. The network was switch-randomized to achieve a range of assortativity values in the same way as for the BA graphs ({\footnotesize see supplementary material Sec. S1 for more details}). For each desired assortativity value, five randomized graphs were generated. The same non-monotonic lifetime–assortativity relationship appears, and the empirical graph is close to the peak of the curve. Assortativity thus emerges as a fundamental structural determinant of message lifetime in nature, and may possess functional significance as discussed below.
\vspace{0.1in}

\section{Discussion}

We have investigated how network structure can directly tune message lifetime under minimal dynamics of message collision and broadcasting. Our findings demonstrate that the lifetime of self-annihilating messages in complex networks is maximized near neutral assortativity. The mechanism is successfully balancing  \emph{amplification}, promoted by assortative hub–hub connections, and \emph{self-annihilation}, where spreading is inhibited by short cycles that proliferate under assortativity and disassortativity. The tradeoff explains why lifetimes are maximized near neutral assortativity and shorten at both ends of the assortativity spectrum. More generally, our results imply that message-message interactions serve in part to decouple dynamics from network architecture.

Our results generalize beyond synthetic networks and have specific relevance in neuroscience. We find that message lifetimes are non-monotonic in randomized connectome-like networks of varying assortativity, and they peak at the neutral assortativity characteristic of the empirical brain network. This suggests that evolved brain structure serves in part to optimize message lifetime in the presence of message interactions. 

Our findings regarding the \textit{mechanisms} of lifetime maximization in networks of neutral assortativity have additional implications for neuroscience. Network architecture suggests that brain networks form \textit{rich clubs} that preferentially link high-degree vertices \cite{van2011rich}; this can be seen as a manifestation of positive assortativity. All else being equal, rich club vertices should thus handle more signal traffic. However, we have shown that the combination of CSA dynamics and neutral assortativity result in message paths that visit vertices in roughly uniform fashion across the graph. Thus, every vertex (region) in the empirical mouse brain network is used with similar frequency. In other words, spreading, annihilating neuronal signals visit brain areas more often than would be predicted by connectivity alone, and activity is not biased in favor of hubs. This finding echoes results from the application of queueing models to the mammal connectome \cite{misic2014communication,misic2014hippocampus} which suggest that message-message interactions can generate information flow that is not expected based on connectivity alone.

Our results resonate with broader theories of brain communication that emphasize limited sustained activity \cite{kaiser2010optimal}. In past work, we have shown that net simulated activity under CSA dynamics (i.e., the average number of vertices containing a message after accounting for collisions) remains low and sparse on the mammal connectome even as injection rate rises \cite{hao2020creative}. Maintaining low and sparsely-distributed activity in the face of changing inputs is a prerequisite for global brain activity \cite{foldiak2003sparse, graham2007}. The CSA model therefore suggests a strategy for achieving sparseness and homeostasis without explicitly-designed mechanisms. We hypothesize that, across species, connectomes typically achieve neutral assortativity, and this can be tested on a wide range of animal connectome data.

We note that, in the brain and elsewhere, collisions can be seen as coincidence detection events. To the extent that multiple vertices can reliably communicate with non-neighbors, a system could achieve selective communication without specific message addressing or timing mechanisms. Because the dynamical rules are the same everywhere, the network structure must contain all the mechanisms for being reliable. This approach may be wasteful (i.e., redundant) but it is a minimal scheme that generates tuneable global behavior. Future studies should investigate how reliably vertex tuples in the brain and other real-world networks can communicate under CSA dynamics and related schemes. Local modifications in particular systems that enable them to harness or suppress collisions for functional needs and to increase efficiency should also be explored. 

Finally, we suggest that our framework may apply quite broadly to engineered communication systems (e.g., blockchains), social networks, epidemic dynamics with exclusion rules, chemical reaction–diffusion processes, and other systems of information spreading where collisions suppress signal lifetime.

\section{Conclusion}

We introduced Copy--Spread--Annihilate Dynamics and our results demonstrated the structural origins of non-monotonic behaviors in spreading processes. We found that degree assortativity is a key driver of message lifetime in networks. Message lifetimes peak at neutral assortativity, a regime where amplification from hub neighborhoods and suppression of annihilation from short cycles strike a balance. This structural tradeoff explains why both highly assortative and highly disassortative architectures curtail lifetime, albeit via different mechanisms. We found that the brain already possesses the network structure needed to maximize lifetime. More generally, by identifying assortativity as a structural control parameter, our work provides a minimal model for situations where dynamics of message broadcasting and interaction can be expected to live or die in a variety of real-world systems. 

\bibliographystyle{apsrev4-2}
\bibliography{refs2}

\clearpage
\onecolumngrid

\section{Supplementary Material}
\setcounter{page}{1}
\setcounter{figure}{0}
\renewcommand{\thefigure}{S\arabic{figure}}
\setcounter{table}{0}
\renewcommand{\thetable}{S\arabic{table}}
\renewcommand{\thesection}{S\arabic{section}}
\renewcommand{\thesubsection}{S\arabic{subsection}}



\section{S1. simulation details}
\begin{figure}[h]
    \centering
    \includegraphics[width=0.7\linewidth]{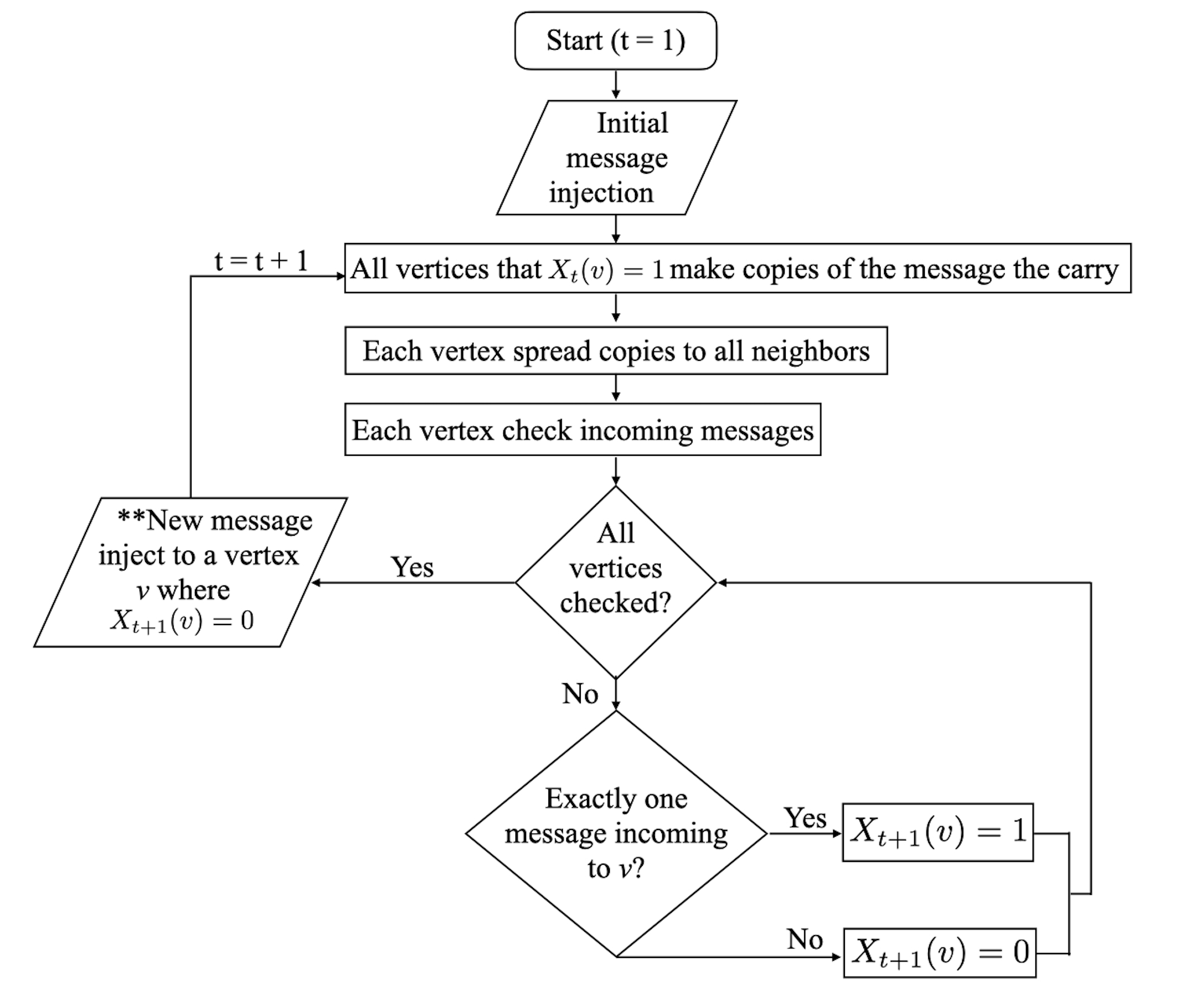} 
    \caption{Numerical implementation of the Copy–Spread–Annihilate dynamics summarized in a flow chart.}
    \label{fig:chart1}
\end{figure}

The numerical simulations involved in this study can be implemented following the flowchart in Fig. \ref{fig:chart1}. For the sequential injection case, if one message is injected at each time step, Fig. \ref{fig:chart1} should be followed exactly. To inject multiple messages as in Fig. \ref{fig:LifeTime2}(b), use multiple messages in the ** step. To simulate the single injection case, skip the ** step. During our study, simulations are stopped when a predetermined number of steps are reached (2000 or 5000 steps), or when all messages are annihilated.

To produce Fig. \ref{fig:LifeTime1}, 34 different $r$ values equally spaced between -0.85 and 0.8 were required. Starting with a 100-vertex graph generated using the Barabási-Albert (BA) algorithm, we employ a randomization process using the degree-based edge-switching algorithm described in \cite{trusina2004hierarchy} to gradually increase or decrease the assortativity until the desired $r$ value is achieved. As a result, a set of 34 daughter graphs are generated using this approach featuring the same degree sequence. Five different sets of graphs (a total of 170) were used to estimate the mean and standard deviation of lifetime curves in Fig. \ref{fig:LifeTime1}. The simulations of message spreading on each set of graphs as described in Fig. \ref{fig:chart1} above were repeated five times. The distances of longest walks traversed by each message and the number of walks that achieved the longest distances are recorded.

To produce Fig. \ref{fig:NodeUsage}, graphs that have assortativites fall in the intervals $[-0.850,-0.65], [-0.1,0.1]$, and $[0.6,0.8]$ are chosen to represent highly assorted, neutral, and highly disassorted graphs, respectively. The simulation of message spreading is again carried out on all five sets of graphs, recording every vertex sequence traversed by all messages through their lifetime. Each simulation lasts 2000 time steps, and the first 100 steps are discarded as transients. The histograms show the frequency at which each vertex participates in walks that are above the 90th percentile in length. The vertices are sorted by degree from low to high.

For Fig. \ref{fig:mouse}, we applied the randomization protocol described above to the empirical mouse connectome repeatedly to generate five sets of 17 daughter graphs whose assortativity values are equally spaced between -0.4 and 0.4.
\newpage

\section{S2. Example path from messages that achieved longer lifetimes}\label{SequentialInj}
\begin{figure}[h]
    \centering
    \includegraphics[width=0.58\linewidth]{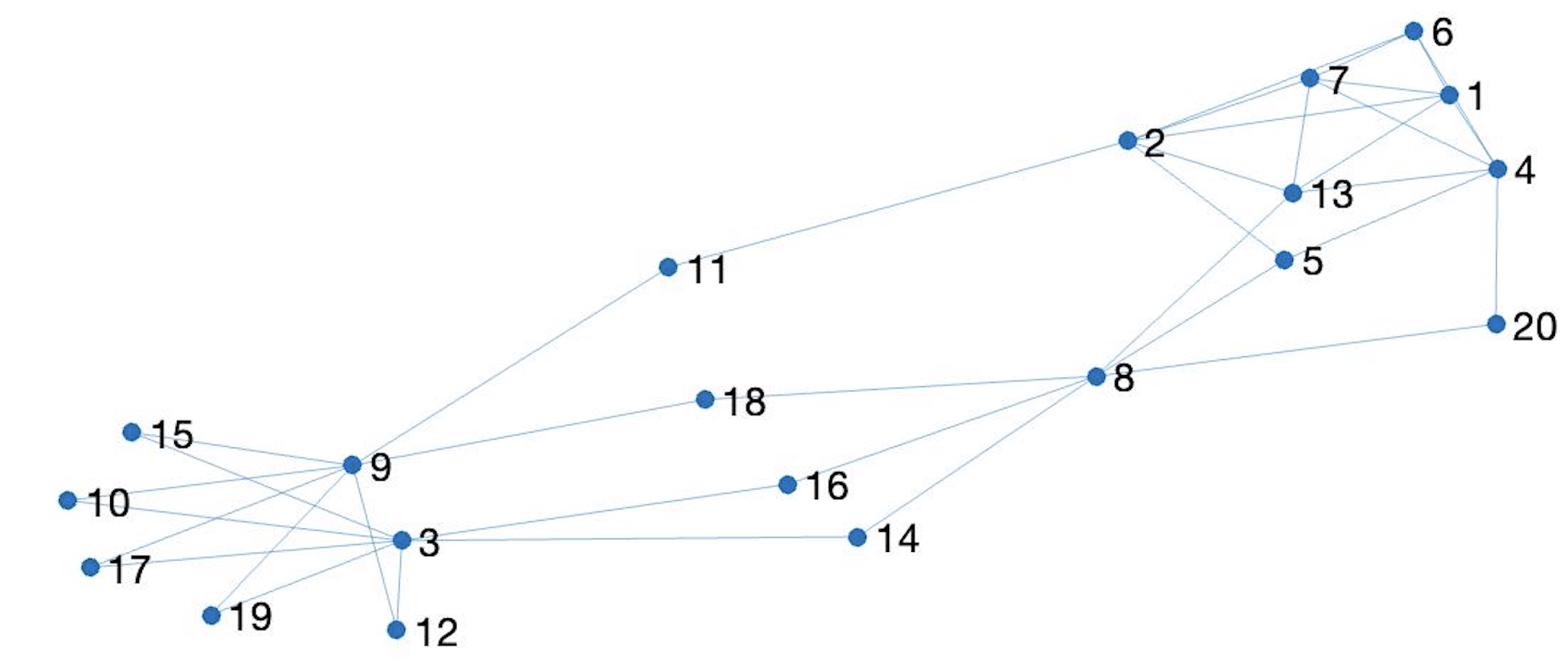}
    \includegraphics[width=0.76\linewidth]{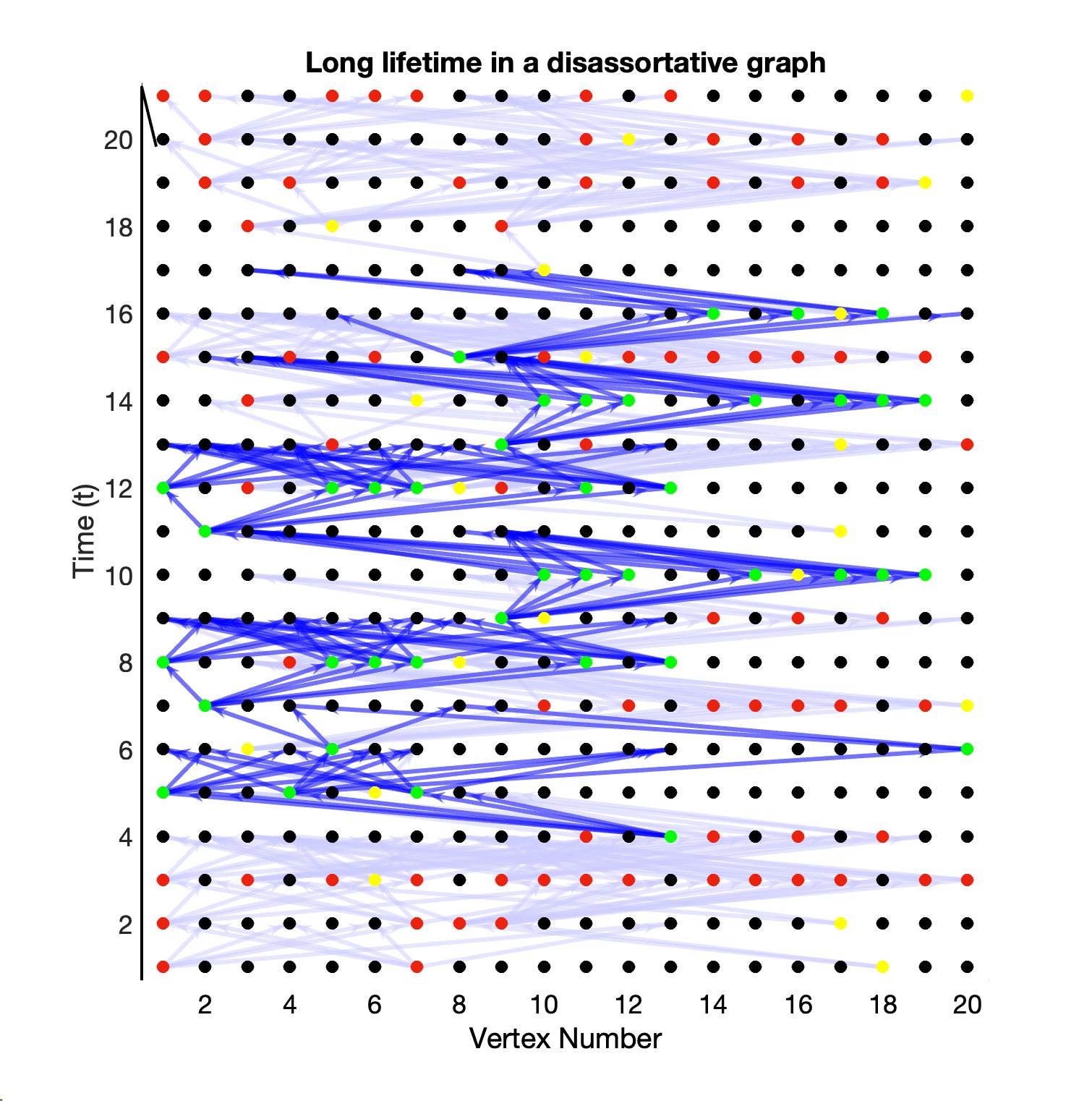}
    
    \caption{Example walks created by one message that achieved a long lifetime on a small disassortative graph ($n = 20, m= 2, r\approx -0.869$, top panel). Green indicates the vertex is carrying the example message. Red indicates the vertex is carrying a different message. Yellow indicates the vertex is receiving the new message that is injected to the system at each step. Black indicates the vertex is not carrying a message. Solid blue arrows show the spreading of the example message, while light blue arrows show the spreading of other messages.}
    \label{fig:PathLow}
\end{figure}

\begin{figure}[h]
    \centering
\includegraphics[width=0.58\linewidth]{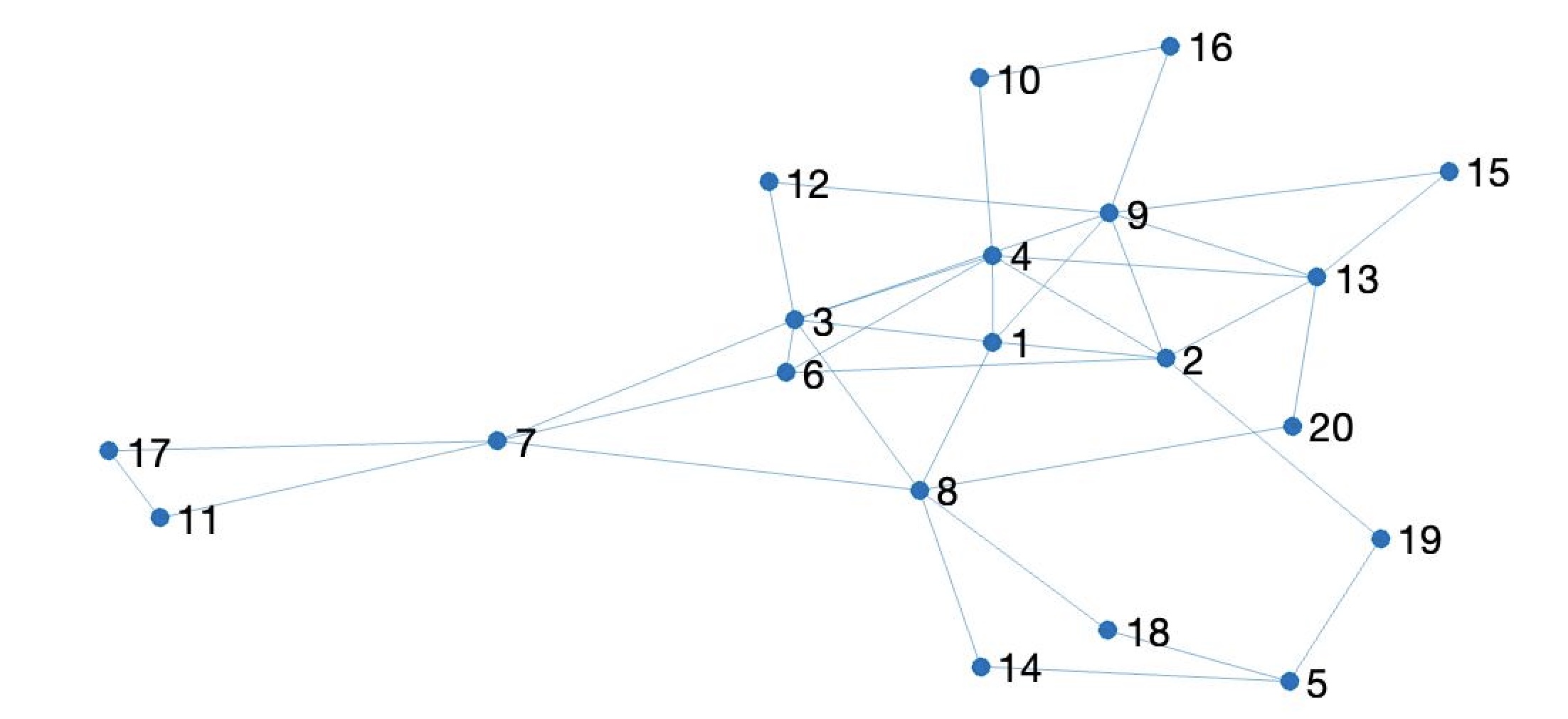}
    \vspace{0.2in}
    \includegraphics[width=0.76\linewidth]{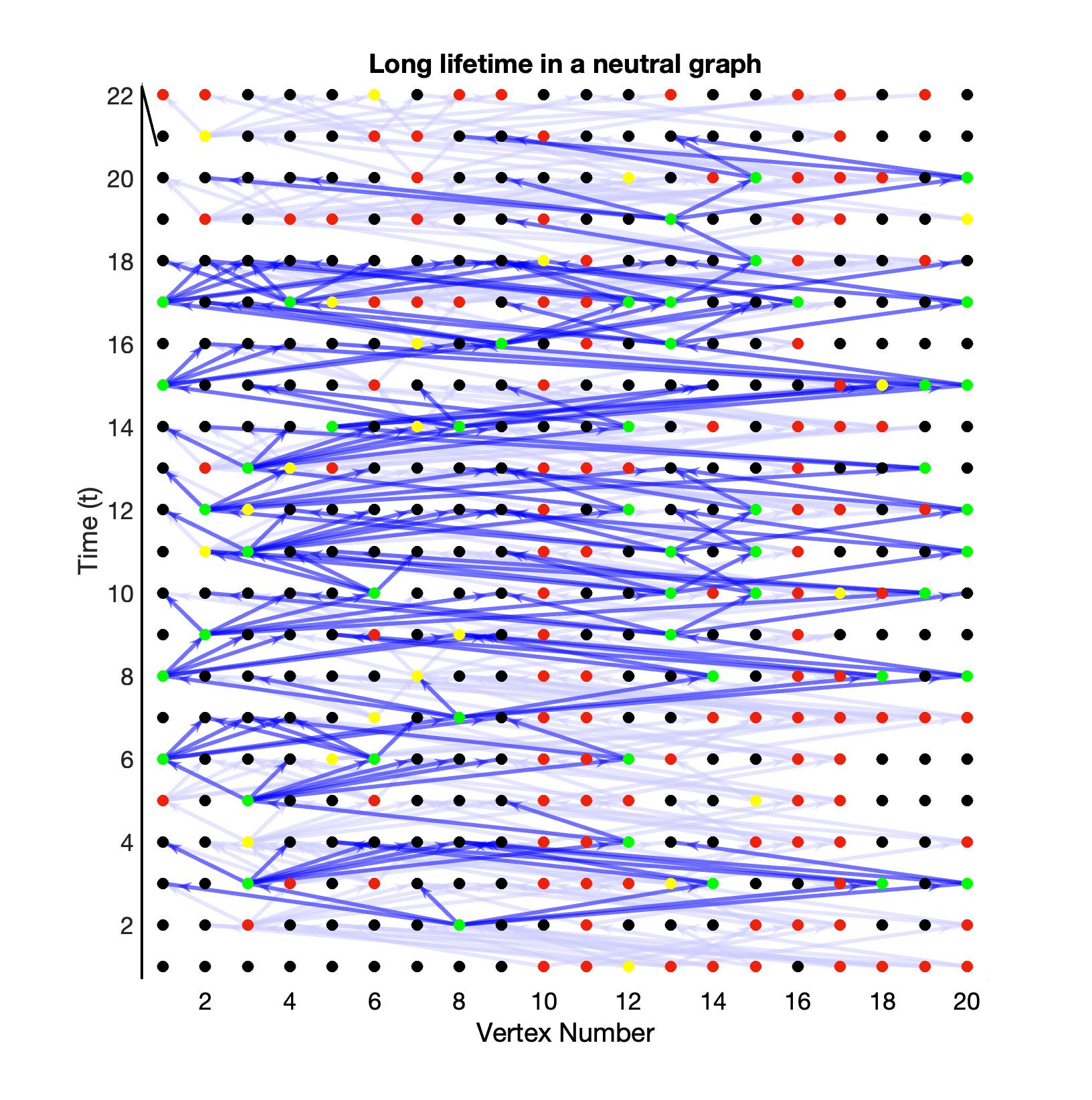}
    \caption{Example walks created by one message that achieved a longer lifetime on a small graph  ($n=20, m=2, r\approx 8.25e-4$, top panel). All color schemes are the same as Fig. \ref{fig:PathLow}}.
    \label{fig:PathMid}
\end{figure}

\begin{figure}
    \centering
    \includegraphics[width=0.58\linewidth]{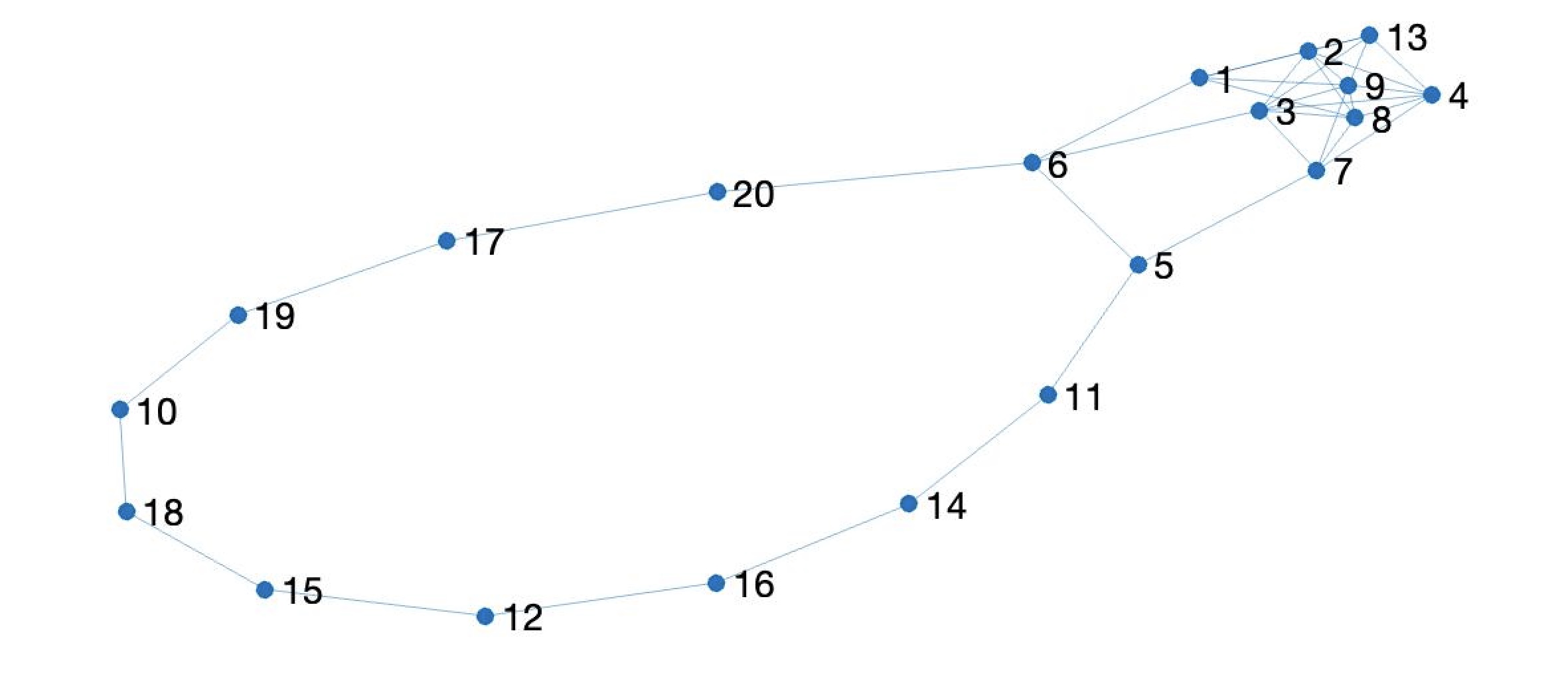}
    \includegraphics[width=0.76\linewidth]{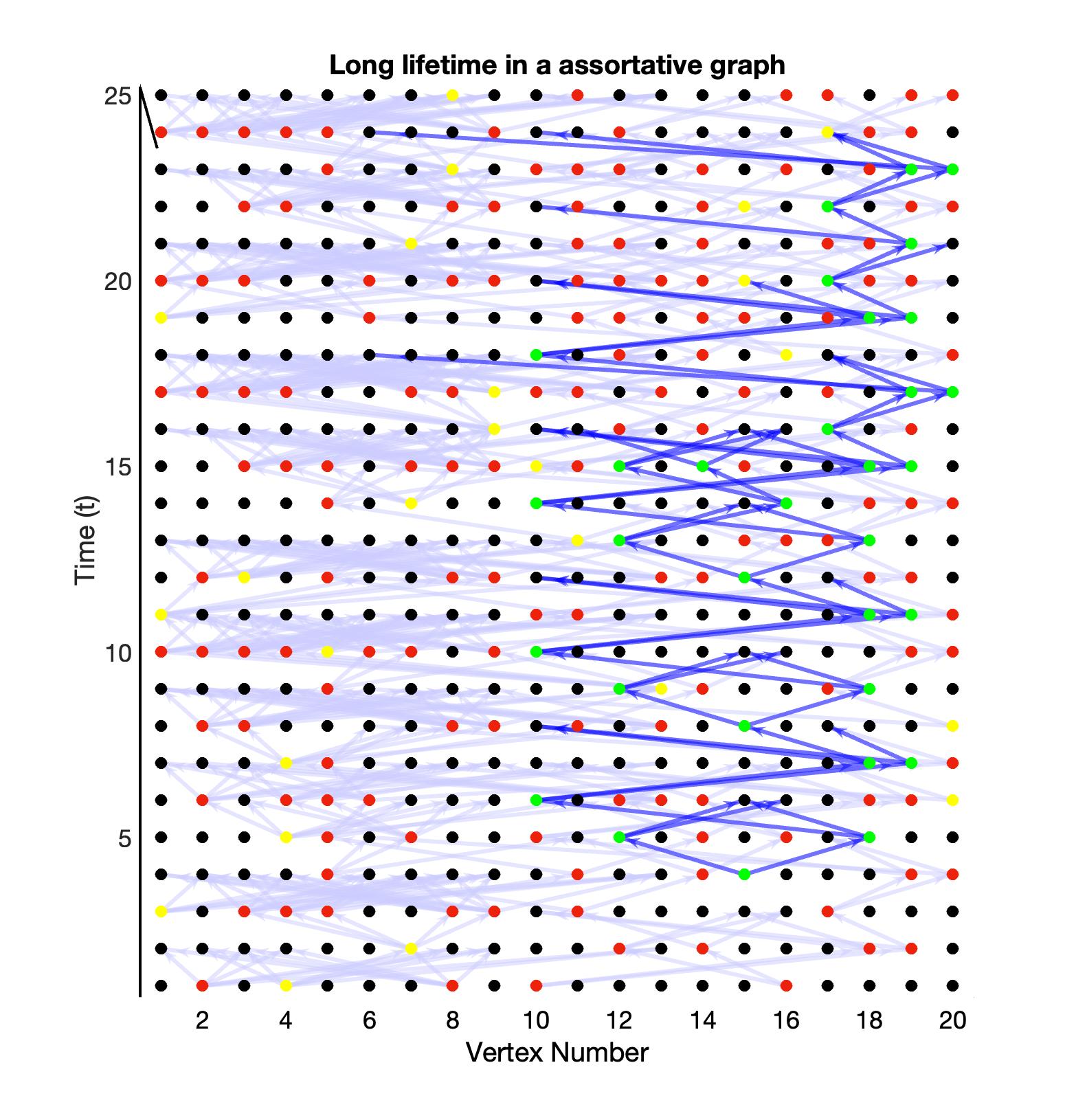}
    \caption{Example walks created by one message that achieved a longer life time on a small assortative graph (n = 20, m=2, $r\approx 0.802$, top panel). All color schemes are the same as Fig. \ref{fig:PathLow}}
    \label{fig:PathHigh}
\end{figure}

In Fig. \ref{fig:scheme}, the copy, spread, and annihilation of one message was illustrated (single injection case). The single injection case is further analyzed in Section S4. All other figures in the main text were results from experiments where, at each timestep, one new message is injected to the system (sequential injection). Here in figures \ref{fig:PathLow}, \ref{fig:PathMid},\ref{fig:PathHigh} we illustrate the characteristics of different graph assortativities, and the space-time diagram of messages in the sequential injection case on those graphs.

At neutral assortativity (Fig. \ref{fig:PathMid}), message lifetime is longer than in Figures \ref{fig:PathLow} and \ref{fig:PathHigh}. The observation that longer-lived messages tend to pass through vertices with particular degree characteristics in different assortativity regimes is also illustrated. These figures show that even when the networks are small (20 vertices) there are already clear differences similar to those shown statistically in Fig. \ref{fig:NodeUsage} for larger graphs.
\\
\\
\\
\\

\newpage
\section{S3. Conditions for non-monotonicity}\label{nonmono}
While searching for the conditions for non-monotonicity, a number of different parameters were evaluated.
\subsection{Graph size does not affect modality}
After noticing graph density does not significantly affect the unimodal shape of the lifetime curves, a variety of graph sizes ($n = 20, 50, 100$) were tested. None of the graph sizes tested show different modality. Results from $n = 50$ are shown in Fig. \ref{fig:LifeTime2}(a).
\subsection{Traffic density does not affect modality}
\begin{figure}[h]
    \centering
    \includegraphics[width=0.48\linewidth]{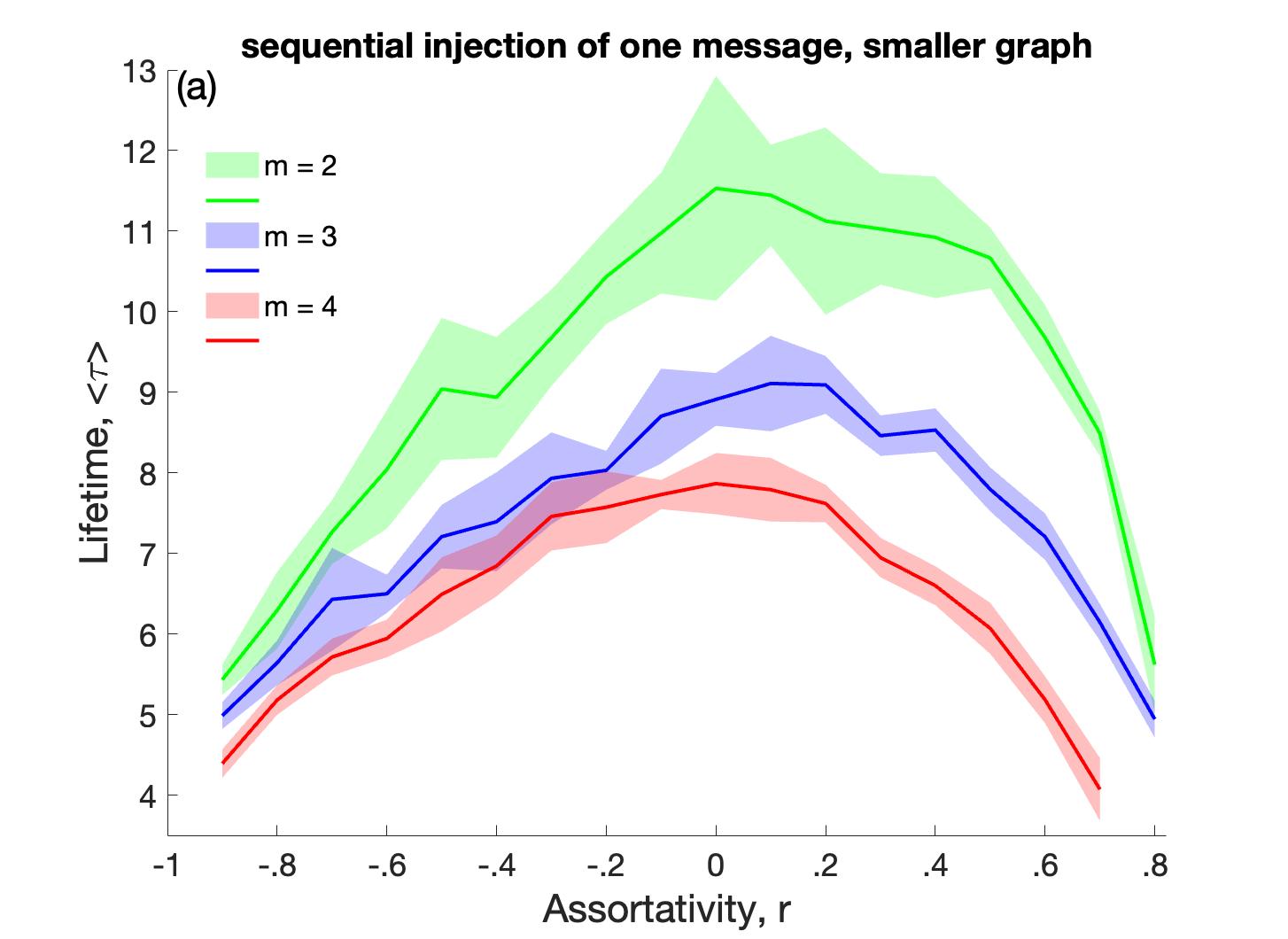}
    \includegraphics[width=0.48\linewidth]{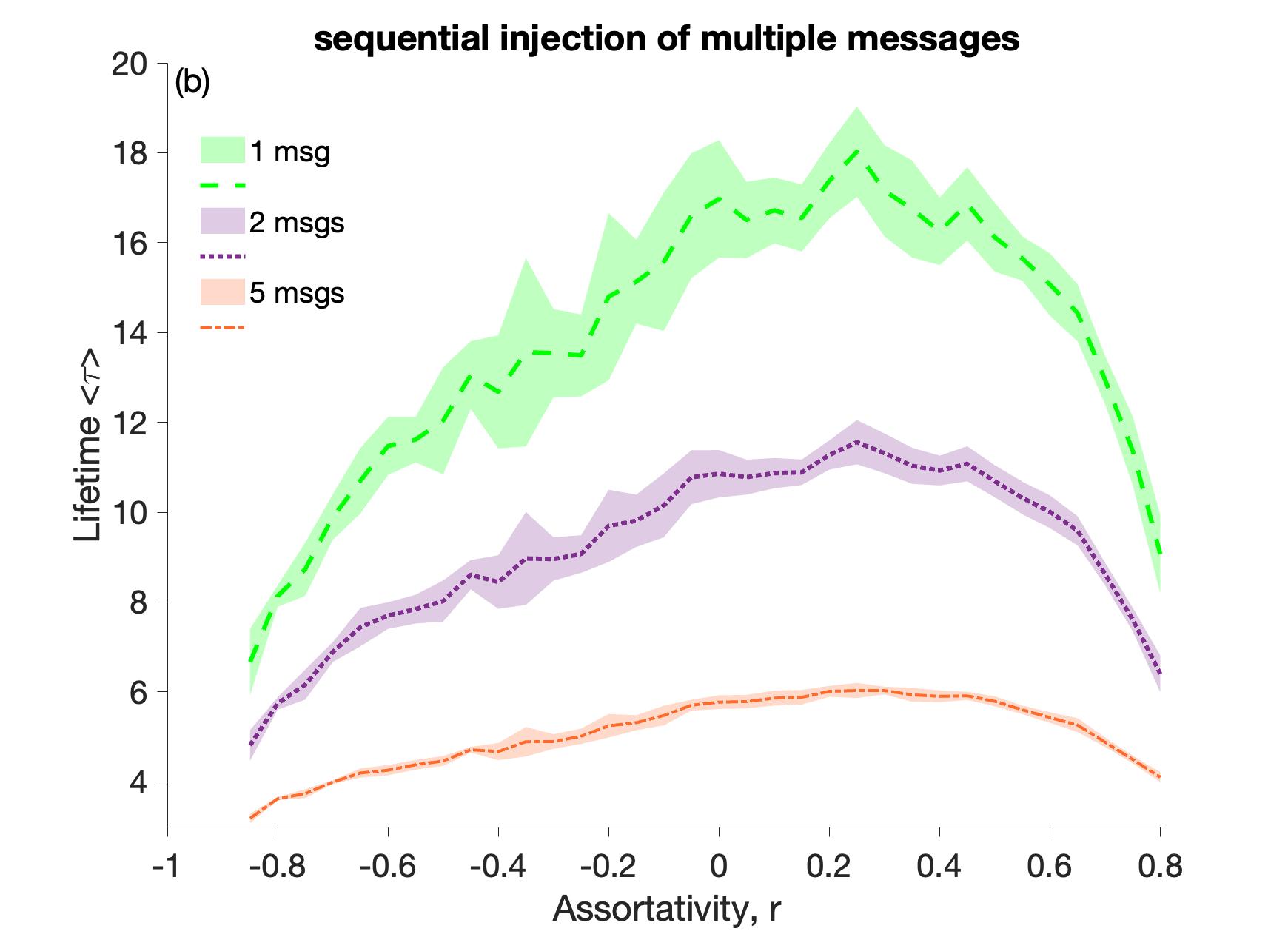}
    \caption{(a) Average message lifetime on a smaller ($n = 50$) BA network using different
parameters of attachment ($m = 2, 3, 4$). One message is injected to these networks at each time step.  (b) Average message lifetime when multiple messages are injected to larger ($n = 100$) BA networks using m = 2 as the parameter of attachment. Each curve shows the mean of five different BA graphs generated using same parameters. Shaded areas indicate standard deviations.
The spectrum of assortativities is achieved using the same degree-based edge-switching algorithm as Fig. \ref{fig:LifeTime1}. }
    \label{fig:LifeTime2}
\end{figure}

A variety of injection rates were also tested to examine whether traffic on the graphs affects the shape of the lifetime curves. As shown in  \ref{fig:LifeTime2}(b), using the same BA graphs ($n = 100, m = 2$) that generated green curves in Fig. \ref{fig:LifeTime1}, one or two or five messages were injected at each time step. Although more messages entering the graph reduces message lifetime in general, the shapes of the curves remains non-monotonic, and peaks near neutral assortativity.
\newpage

\section{S4. Vertex-participation histograms and structural strategies}
\begin{figure}[h]
    \centering
    \includegraphics[width=0.9\linewidth]{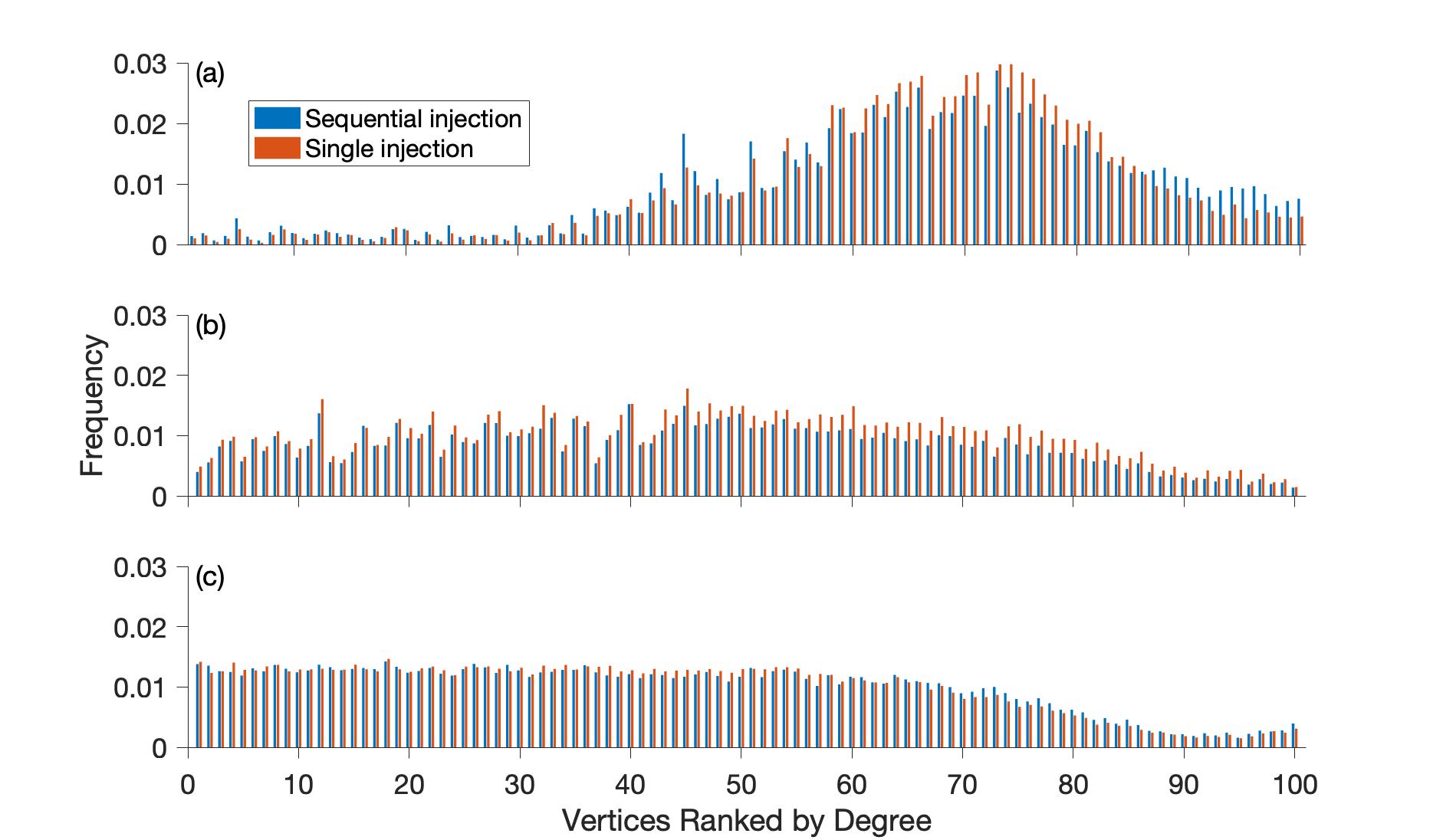}
    \caption{Vertex involvement comparison: We examine the single injection case, where one initial message is injected to a network but no further messages are injected, in comparison to the sequential injection method. Results are collected from five different networks of 100 vertices generated using the Barabási–Albert (BA) algorithm and the parameter of attachment $m = 3$. Vertices are sorted by degree from low to high. (a) shows in disassortative graphs ($-0.85 < r < -0.65$) higher degree vertices are frequented by successful walks, (b) shows when assortativities are neutral ($-0.1 < r < 0.1$) vertex involvement is fairly uniform, and (c) shows that, in assorted graphs
($0.6 < r < 0.8$), lower degree vertices are preferred, while high degree vertices are avoided in the longest walks.}
    \label{fig:NoAddHist}
\end{figure}

One may have noticed that annihilation caused by copies from different messages were not discussed in this manuscript. We address the reasons here with Fig. \ref{fig:NoAddHist}. This figure compares vertex involvement between the single injection method and the sequential injection method (one message per time step). In all three regimes of assortativity (negative, neutral, and positive), the preferences in vertex involvement are nearly identical between the two injection methods. This result supports our argument that self-annihilation is one of the primary forces that shapes the lifetime curves. Combining this result with the fact that injecting more messages into the graphs does not change the shape of the lifetime curves (Fig. \ref{fig:LifeTime2}(b)), we are convinced that annihilation caused by copies from different messages is not a primary condition that contributes to the non-monotonic shape of the lifetime graphs. 

\end{document}